\documentclass[twocolumn,showpacs,preprintnumbers,amsmath,amssymb]{revtex4}
\usepackage{graphicx}
\usepackage{epstopdf}

\newcommand{\Bpos}{\vec{B}(\vec{x})}

\newcommand{\Mpos}{\vec{M}(\vec{x})}

\begin{document}
\draft

\title{Hysteresis multicycles in nanomagnet arrays}

\author{J. M. Deutsch, Trieu Mai, and Onuttom Narayan}
\affiliation{
Department of Physics, University of California, Santa Cruz, CA 95064.}

\date{\today}

\begin{abstract}
We predict two new physical effects in arrays of single-domain
nanomagnets by performing simulations using a realistic model
Hamiltonian and physical parameters. First, we find hysteretic
multicycles for such nanomagnets.  The simulation uses continuous
spin dynamics through the Landau-Lifshitz-Gilbert (LLG) equation.
In some regions of parameter space, the probability of finding a
multicycle is as high as $\sim 0.6.$ We find that systems with
larger and more anisotropic nanomagnets tend to display more
multicycles. Our results also demonstrate the importance of disorder
and frustration for multicycle behavior. Second, we show that there
is a fundamental difference between the more realistic vector LLG
equation and scalar models of hysteresis, such as Ising models.  In
the latter case spin and external field inversion symmetry is obeyed,
but in the former it is destroyed by the dynamics, with important
experimental implications.  
\end{abstract}


\maketitle 

\section{Introduction} Hysteresis in magnets~\cite{Steinmetz,Bark}
is a paradigm for all
history dependent behavior in nature~\cite{colloids,cdw,alloys}. In addition, hysteresis is
the cornerstone of the magnetic data storage industry, and of great
technological importance~\cite{techref}. Understanding the full possibilities of
magnetic hysteresis is thus important both for fundamental insights
and practical implications.

Recently, we have shown~\cite{multicycle} that spin glasses can exhibit stable
``multicycle" hysteresis loops, in that when the external magnetic
field is cycled adiabatically, the magnetization returns to itself
after $m > 1$ cycles of the magnetic field. This behavior should
be experimentally observable in spin glass nanoparticles at low
temperature. Thus simple one-cycle hysteresis loops, although
ubiquitous and generally assumed to be universal, are only part of
a much richer phenomenon.

In our previous work~\cite{multicycle}, we used the standard
Edwards-Anderson spin glass Hamiltonian~\cite{sg2}, with Ising spins and nearest
neighbor interactions in three dimensions. Zero temperature dynamics
or Monte Carlo dynamics at low temperature were used. Starting from
saturation, the magnetic field was lowered adiabatically, and then
cycled repeatedly over a suitably chosen range. When the system
reached steady state, the order of the multicycle (as defined in
the previous paragraph) was measured. Whether a multicycle is
present, and if so, its order, depended on the realization of
randomness, varying from one system to another.

There were two weaknesses in the previous work. First, the systems
considered were susceptible to thermal noise meaning that very low
temperatures were needed to prevent fluctuations from destroying
the periodicity. Second, the model considered was an Ising 
model. Although this has well defined equilibrium statistics,
it is not clear that it adequately represents the dynamics.

In this paper, we overcome the problems of the previous paragraph
by examining a different system as a candidate for multicycle
hysteresis behavior: an array of magnetic nanoparticles.  With
current technology, it is standard to fabricate such magnetic arrays
according to specification, and a wide variety has been studied
experimentally~\cite{Schmidt2} and these have important applications
for magnetic storage technology, so called ``patterned media"~\cite{patterned}.  For
our purposes, these systems offer many experimental advantages over
the spin glass nanoparticles that we considered earlier. One can
build ``designer arrays" with optimized parameters that maximize
multicycles and select them in a predictable manner. Spatially
resolved measurements are possible, unlike for the spin glass case,
making experimental observations of multicycles more practical.
The relatively large size and shape of these also greatly reduces
the effects of thermal noise, (which is one important reason that
they are useful in disk drive recording).  Finally, the ability to
address small regions of the array instead of only applying a uniform
external magnetic field, and the sensitivity of multicycle phenomena
to relatively small changes in system parameters, may open up
the possibility of using these systems for computation.

Through numerical simulations, we demonstrate that multicycles can
be seen in an array of pillars made of ferromagnetic material,
coupled to each other through dipolar forces, arranged in square
and triangular lattices. The external magnetic field is applied
perpendicular to the lattice, {\it i.e.\/} parallel to the axes of
the pillars; to be specific, we will refer to this as the vertical
or $z$-direction.  The existence of multicycles is robust, persisting
over a large range of system parameters.

We also examine the importance of frustration and disorder in
achieving multicycles, a question that has been of great interest
in spin glass research~\cite{sg2}. We find that when a square lattice is used
instead of a triangular one, multicycles are not as likely to be
seen. As discussed later in this paper, the magnetization in the
pillars prefers to be approximately vertical, so that the dipolar
coupling between the pillars is antiferromagnetic. For a square
lattice, the dipolar forces between nearest neighbors are not
frustrated, unlike the case for a triangular lattice. (There is 
still some frustration because of further neighbor interactions.)
This result emphasizes the importance of frustration.

For the spin glass system considered earlier, disorder was explicitly
present through the random bond strengths. For the nanoparticle
array, although the bond strengths are not random (unless the
spacings between the pillars are varied), there is crystal anisotropy,
arising from the fact that the magnetization prefers to align itself
in a specific direction relative to the crystal axes. Because of
the way in which the pillars are grown, the orientation of the
crystal axes is different in each pillar, and random. Even if the
crystal anisotropy energy is small compared to the dipolar (and
other) energies, we find that it is sufficient to cause multicycles
in an otherwise regular triangular lattice. The order of the
multicycle for a specific sample depends on the orientation of the
crystalline axes in its pillars. Although in this paper we only
consider the case of randomly oriented crystal axes, if the pillars
could be grown with the orientations specified, it would be possible
to make arrays whose hysteresis loop are multicycles of desired
order.

Regardless of the source, it is desirable to have {\it some\/}
inhomogeneity in the model to see multicycles. Otherwise, when the
external magnetic field is reduced from saturation, the magnetic
moments would be stuck in unstable equilibrium; in practice, thermal
noise would push the system out of this unstable point. Although
this effect would not be so extreme subsequently when the external
field is cycled over a range that does not reach saturation, it
is clear that to reduce the sensitivity to thermal noise, and thereby
increase definite multicycle periodicity, explicit quenched disorder
is desirable.

The model used for the dynamics in our numerical simulations is
discussed in detail in the next section; the magnetic moment of
each pillar is treated as a single `Heisenberg spin', {\it i.e.\/}
with its orientation as a continuous variable, with continuous time
dynamics.  This is in contrast to the earlier spin glass work, with
Ising spins and discrete (event driven) dynamics. Even though the
dynamics are continuous, as discussed in the next section there is
a shape anisotropy energy for tall pillars that causes the magnetization
to be nearly vertical, and jump from up to down (or vice versa) as
the external field is changed. This jump can trigger instabilities
in other pillars, forming an avalanche. We believe that in order
for multicycles to be seen, it is essential for the interaction
between pillars to be sufficiently strong to cause avalanches; in
the extreme case, when the pillars are independent, it is clear
that a one-cycle hysteresis loop would be seen. However, avalanches
are not sufficient to produce disorder: for disordered nearest
neighbor Ising ferromagnets, the phenomenon of Return Point Memory~\cite{Sethna,RPM} (RPM) can
be proved, precluding multicycles. This emphasizes the need for
frustration. 

This is the first paper, as far as we are aware, that studies
adiabatic hysteresis loops
in magnetic systems using the more fundamental
LLG equations rather than simplified relaxational dynamics. Experience
from critical phenomena might lead one to believe that the difference
between this approach and previous work would be trivial. However
there is an important physical difference that we believe has been
overlooked. LLG dynamics destroy symmetry under global spin flip,
even though the Hamiltonian is symmetric under this operation. This
result, which will be discussed further in the next section of this
paper, has significant experimental implications~\cite{Sorensen9}.

In the next section of this paper, the dynamical equation used in
the numerical simulations is introduced, and various terms in the
model Hamiltonian are calculated. Details of the numerics are given
in Section III, and the results thereof are presented in Section
IV.

\section{Classical Spin Dynamics and the model Hamiltonian}
Microscopically, the evolution of classical spins is described by
the Landau-Lifshitz-Gilbert (LLG) equation of motion~\cite{LLGref}.  The LLG
equation is the simplest equation describing micromagnetic dynamics
which contains a reactive term and a dissipative term:
\begin{equation}
\frac{d\vec{s}}{dt} = -\gamma_1 \vec{s} \times \vec{B} -\gamma_2 \vec{s} 
\times (\vec{s} \times \vec{B}), 
\label{LLG}
\end{equation} 
where $\vec{s}$ is a microscopic spin, $\vec{B}$ is the local
effective field, $\gamma_1$ is a precession coefficient, and
$\gamma_2$ is a damping coefficient.  The effective field is $\vec{B}
= - \frac{\partial \mathcal{H}}{\partial \vec{s}}+\vec{\zeta}$,
where $\mathcal{H}$ is the Hamiltonian and $\vec{\zeta}$ represents
the effect of thermal noise.  Terms in the Hamiltonian will be
discussed and computed later in this section.

The reactive term of the LLG equation describes the precession of
the spin about its local field, with the angle between the two
remaining constant.  (The coefficient of the reactive term $\gamma_1$
will be set to unity throughout this paper unless otherwise noted.)
The dissipative term aligns the spin with its local effective field.
The cross products in the dissipative term ensure that only the
tangential component of the field causes damping, since the length
of $\vec{s}$ cannot change. The relaxation time is inversely related
to the damping coefficient $\gamma_2$.  Reasonable approximations
for $\gamma_2$ are difficult to obtain, but it will be shown that
the hysteresis multicycle phenomenon studied in this paper is present
for a large range of $\gamma_2$.

With current technology, nanomagnetic pillars that are approximately
50nm wide and 100nm tall can be made of ferromagnetic materials
such as nickel~\cite{Schmidt2}.  For such small pillars, it is found
that the ferromagnetic coupling between the atoms dominates the
antiferromagnetic dipolar interactions.  Thus the entire pillar
consists of a single magnetic domain.  Using the lattice constant
of nickel, each pillar holds approximately $10^7$ atoms, allowing
us to treat the pillar as a continuous magnetic medium.  Edge effects
such as splaying near the boundaries are neglected, and the pillar
is treated as a saturated nanomagnet with uniform magnetization.
Each single-domain nanomagnet can be viewed as a single degree
of freedom: a magnetic moment of fixed magnitude, whose orientation
represents the direction of the magnetization~\cite{Stoner}.  The time evolution
of this magnetic moment has the \textit{same} structure as the
micromagnetic LLG equation: a reactive part, $-\gamma_1 \vec{s}
\times \vec{B}$, and a dissipative part, $-\gamma_2 \vec{s} \times
(\vec{s} \times \vec{B})$, although $\gamma_2$ is different from
its microscopic value.  (Henceforth $\vec{s}$ will denote a unit
vector in the direction of the magnetic moment of a pillar, rather
than an individual spin.)  As before, the field $\vec{B}$ is given
by $- \frac{\partial \mathcal{H}}{\partial \vec{s}}$; the large
number of spins evolving in unison in each pillar allows the thermal
noise $\vec{\zeta}$ to be neglected.

To complete the specification of the dynamics of the magnetic moments
through Eq.(\ref{LLG}) , the Hamiltonian has to be calculated in
terms of the magnetic moment of each pillar.  The various terms in
the Hamiltonian are discussed in the following paragraphs.

First, the geometry of the pillars introduces a shape anisotropy
term in the Hamiltonian:
\begin{equation}
\mathcal{H}_{SA} = -d_{z} \displaystyle\sum_{i}s_{z,i}^{2},
\label{shape}
\end{equation}
where $d_{z}$ is a constant to be calculated in the next paragraph
and $s_{z,i}$ is the z-component of the magnetic moment of the
$i^{th}$ pillar.  Shape anisotropy energy is present because of the
dipolar interactions between the individual spins within a pillar.
Qualitatively, if the magnetization of a tall skinny pillar is
vertical, the spins are predominantly lined up ``head to toe".
This configuration has a lower energy than when the magnetization
is horizontal, in which case the spins are predominantly side by
side.  For a short wide disk, the effect is clearly reversed.

For the case of tall pillars, the shape anisotropy reduces the
magnetic moment to an almost Ising like variable, that can
(approximately) only point up or down.  The dynamics of anisotropic
and isotropic spins are qualitatively different, with avalanche
phenomena more likely to occur in the former than the latter.  As
mentioned earlier, we believe that avalanches are necessary for
hysteretic multicycles.  Note that even when the shape anisotropy
is large, we evolve each magnetic moment according to Eq.(\ref{LLG})
rather than as an Ising variable, {\it i.e.\/} with an orientation
that evolves continuously with time, though the shape anisotropy
causes rapid transitions from up to down states.

Deriving the form of $\mathcal{H}_{SA}$ and the value of $d_{z}$
requires solving a magnetostatic problem. The energy of the field
due to the microscopic spins in a single pillar is $W = {1\over
2\mu_{0}}\int d^{3}x |\Bpos|^{2},$ where $\Bpos$ is the magnetic
field at $\vec x$ due to the spins.  Through Ampere's law and vector
calculus manipulations, we can rewrite this in terms of the
magnetization, $\Mpos.$ For uniform magnetization, the result can
(up to an additive constant) be converted to a surface integral,
similar to electrostatics, with the self-energy of the magnetic
surface ``charge" to be calculated.  For cylindrical pillars, which
we consider in the rest of this paper, if the magnetization has
magnitude $M_0$ and makes an angle $\theta$ to the vertical, and
$R$ and $h$ are the radius and height of the pillars, the final
result is
\begin{equation}
W = \mu_0 M_0^2 R^3 F(\frac{h}{R}) \cos^{2}{\theta}
\end{equation} 
up to an additive constant independent of $\theta$.  Comparing with
Eq.(\ref{shape}), we see that $d_z = -\mu_0 M_0^2 R^3 F(h/R)$, where
$F(h/R)$ is a function of the aspect ratio that can be evaluated
numerically.  With $M_0$ equal to the saturation magnetization for
nickel, $4.84\times10^5 A/m$~\cite{Schmidt3}, the values of $d_z$
for different sized pillars are given in Table~\ref{table1}.  If
the pillars are ellipsoidal instead of cylindrical, $d_z$ can be
obtained analytically instead of numerically~\cite{Stoner,Schmidt3}.

\begin{table}
\caption{\label{table1}Calculated shape anisotropy coefficients
$d_z$ for pillars with radius $R$=30 nm and different aspect ratios
$h/R$. The ratios between crystal anisotropy coefficients and $d_z$
are also given.}
\begin{ruledtabular}
\begin{tabular}{cccc}
 $h/R$ & $d_{z}$ ($\times 10^{-18}$J) & $|K_{1}/d_{z}|$ &
 $|K_{2}/d_{z}|$\\
\hline
0.5 & -2.866 & 0.0740 & 0.0296 \\
1 & -2.689 & 0.1577 & 0.0631 \\
2 & 0.769 & 1.102 & 0.441 \\
3 & 5.754 & 0.2211 & 0.0885 \\
4 & 11.30 & 0.1501 & 0.0600\\
5 & 17.11 & 0.1240 & 0.0496 \\
10 & 47.41 & 0.0894 & 0.0358 \\
\end{tabular}
\end{ruledtabular}
\end{table}

A second form of anisotropy energy is caused by the crystal structure
of Ni. As mentioned earlier, the crystal axes give a preferential
direction to the magnetization, independent of the shape of the
pillar.  Nickel has a face-centered cubic structure which tends to
align spins in the [111] direction.  If $\alpha_{x,i}$, $\alpha_{y,i}$,
and $\alpha_{z,i}$ are the direction cosines of the magnetization
of the $i^{th}$ pillar to its $(x,y,z)$ crystal axes, the crystal
anisotropy energy can be expanded in powers of $\alpha$.  The first
two terms are~\cite{Stoner,Schmidt3}
\begin{eqnarray}
\mathcal{H}_{CA} = \displaystyle \sum_{i} 
-\frac{K_{1}}{2}(\alpha_{x,i}^{4}+\alpha_{y,i}^{4}+\alpha_{z,i}^{4})
+K_{2}\alpha_{x,i}^{2}\alpha_{y,i}^{2}\alpha_{z,i}^{2},
\end{eqnarray}
where an additive constant has been dropped. The material parameters
$K_1$ and $K_2$ can be obtained by multiplying the experimentally
obtained energy densities with the volume of the pillars.  Crystal
anisotropy energy densities have approximate values of $-5\times10^{3}
J/m^3$ and $-2\times10^{3} J/m^{3}$ for $K_1/vol$ and $K_2/vol$
respectively~\cite{Schmidt3}.  Though the ratio of $K_2/K_1$ is
$\sim$ 0.4, the first term dominates, since it has 2 less powers
of $\alpha$.

In addition to the shape and crystal anisotropies that affect each
pillar by itself, there is dipolar coupling between pillars.
Microscopically, this is similar to the shape anisotropy energy,
except that it arises from interactions between spins on different
pillars.  The resultant interaction energy is of the form
\begin{eqnarray}
\mathcal{H}_{dip} = \displaystyle\sum_{i,j\neq i} \vec{s}_i 
\cdot A({\vec{r}_{ij}})\cdot \vec{s}_j.
\end{eqnarray}
$A({\vec{r}_{ij}})$ is a second rank tensor that depends only on
the separation of the pillars.  The elements of $A({\vec{r}_{ij}})$
are determined by numerically solving integrals similar
to the integrals for the shape anisotropy energy.

The last term in the Hamiltonian is due to the external magnetic
field, which we take to be in the z-direction.  The form of this
term is the conventional one, $\mathcal{H}_{ext} = -B_e\displaystyle\sum
s_{z,i}$.  Hysteresis occurs as $B_e$ is varied adiabatically, with
the system evolving according to the LLG equation.

In summary, the full Hamiltonian has four terms: shape anisotropy,
crystal anisotropy, external field, and dipole-dipole interaction.
The first two are properties of the pillars individually, the
external field term is the term that is adiabatically changed to
observe hysteresis, and the dipole term is an interaction between
pillars:
\begin{widetext}
\begin{eqnarray}
\mathcal{H}=\displaystyle\sum_{i} 
[-d_{z} s_{z,i}^2 -\frac{K_{1}}{2}(\alpha_{x,i}^{4}+
\alpha_{y,i}^{4}+\alpha_{z,i}^{4}) 
+K_{2}\alpha_{x,i}^{2}\alpha_{y,i}^{2}\alpha_{z,i}^{2}
- B_e s_{z,i}+ \displaystyle\sum_{j\neq i} 
\vec{s}_i \cdot A({\vec{r}_{ij}})\cdot \vec{s}_j].
\label{hamilt}
\end{eqnarray}
\end{widetext}

In the numerics, the actual values of the coefficients in the
Hamiltonian are inconsequential, and only the ratios of terms are
relevant.  Table~\ref{table1} shows the results of the calculations
described above for $d_z$, $K_1$, and $K_2$.  Evidently, for the
dimensions of the nanomagnetic pillars of interest, the shape
anisotropy term is larger than the crystal anisotropy.  The dipolar
coupling is also small compared to $d_z$ for the lattice spacings
of interest.  In the simulations, all energies are normalized to
$d_{z}$.  Although $d_{z}$ is dominant for pillars with aspect
ratios of interest, the other terms must be included in the Hamiltonian
because they affect the dynamics qualitatively.  Without the dipole
term, each pillar would be isolated.  The crystal anisotropy term
introduces quenched randomness in the system, and determines the
order in which the $s_i$'s flip when the magnetic field is changed;
its importance has been discussed towards the end of Section I.
Thus the Hamiltonian of Eq.(\ref{hamilt}) has all the important
terms that have to be kept.

As mentioned in Section I, although the Hamiltonian of Eq.(\ref{hamilt})
is invariant if all the $\vec{s}_i$'s are flipped (along with the
external magnetic field), the dynamics of Eq.(\ref{LLG}) are not.
In Eq.(\ref{LLG}), the left hand side and the dissipative term on
the right hand side change sign, but the reactive term does not.
Therefore the spin inversion symmetry, although relevant to equilibrium
static properties, does not apply to the non equilibrium dynamics
appropriate for hysteresis.  In particular, the two sides of the
major hysteresis loop are not complementary to each other.

\section{Numerics} 
As mentioned earlier, the pillars are modeled as single degrees of
freedom which follow the LLG equation of motion.  The effective
field for each magnetic moment in the LLG equation is the `spin'
derivative of the Hamiltonian of the previous section.  Pillars are
placed on a 2-d triangular lattice, to maximize the frustration of
the dipolar bonds. All systems studied are $4\times 4$ lattices with open
boundary conditions.  The positions of the pillars are $i \hat x +
j (\hat x/2 + \hat y\sqrt 3/2)$ with $i, j = 0,\ldots,3.$ The
orientation of the crystallographic axes is separately and randomly
chosen for each pillar.  Depending on the choice of these random
orientations, a sample can have multicycles of various orders $m$,
or a simple hysteresis loop ({\it i.e.\/} $m=1$).  Square lattices
are also considered.

The dimensions of the cylinders and the separation between them,
the external field range, and the damping coefficient $\gamma_2$
are input parameters.  These are used to calculate $d_{z}$, $K_{1}$,
$K_{2}$, and the elements of $A({\vec{r}_{ij}})$.  All the input
parameters can be adjusted to maximize the occurrence of multicycles,
except $\gamma_2$.  Since $\gamma_2$ is a property of the material,
but is unknown, we make sure that the results reported here are
valid over a wide range of $\gamma_2$.

Numerical modeling of the adiabatic field variation is straightforward.
The external field is lowered or raised by a small field step,
$\delta B_e$.  To optimize speed, the field step $\delta B_e$ is
adjusted adaptively, since a small step is required during avalanches.
The effective field is then calculated and the system evolves by a
small time step, $\delta t$, with this field.  This time evolution
is repeated, without changing $B_e.$ Numerical integration of the
LLG equation is implemented using the fourth order Runge-Kutta
algorithm.  Once the system ``settles" to a stationary state,  the
external field is changed again.  Waiting for the system to reach
a stationary state is equivalent to varying $B_e$ slower than all
the dynamics of the system, {\it i.e.\/} adiabatically.

The requirement for settling is that the configuration after the
evolution by a time step $\delta t$ is essentially the same as the
configuration before.  In practice, some numerical tolerance is
allowed, and the initial and final configurations must differ by
less than this tolerance.  The sums $\delta s_{x,i}^2 + \delta s_{y,i}^2
+\delta s_{z, i}^2$ for each $i$ and $\sum_i\delta s_{z,i}$ must 
all be less than $10^{-11}$ in one time step
for the system to be considered stationary. The results reported
here are insensitive to a reduction of $\delta B_e$, $\delta t$,
or the tolerance, and therefore represent adiabatic field variation
with continuous time dynamics.

Starting from a large positive $B_e$ (so that all the pillars are
magnetized upwards) the external field is lowered and cycled
adiabatically over a range $[-B_e^{max},B_e^{max}]$.  The configuration
$\{\vec{s}_i\}$ is compared at $B_e^{max}$ after each cycle.  If
$\{\vec{s}_i\}$ is the same after every $m$ occurrences of $B_e =
B_e^{max}$, the system is in an $m$-cycle.  Similar to the condition
for settling, the configurations match up to a tolerance; we have
verified in numerous cases that the tolerance does not introduce
spurious multicycles.  A tolerance of $10^{-4}$ for each component
of the magnetization was found to be sufficient.  Initially, the
system undergoes a transient period of a few cycles of $B_e$ before
reaching a limit cycle.
\begin{figure}
\begin{center}
\includegraphics[width=3in]{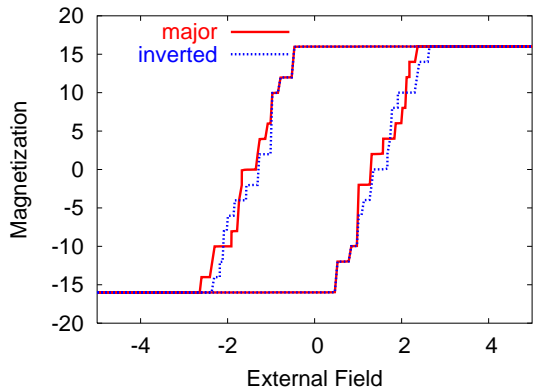}
\caption{Major hysteresis loop (solid curve) for a $4\times 4$ 
triangular lattice of pillars. The steps demonstrate avalanching 
dynamics. In order to see that the two halves of the loop are not 
complementary, the dashed curve shows the same hysteresis loop, 
with $\vec{M}\rightarrow -\vec{M}$ and $\vec{B}\rightarrow 
-\vec{B};$ the solid and dashed curves clearly do not coincide.}
\label{fig2}
\end{center}
\end{figure}

\begin{figure}
\begin{center}
\includegraphics[width=3in]{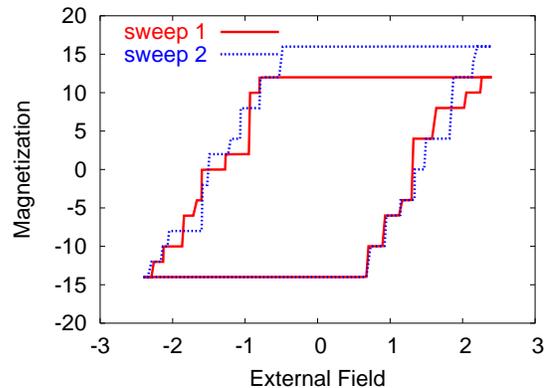}
\caption{A hysteresis 2-cycle, starting at $B_e = -B_e^{max}$.  
The solid curve is a hysteresis
loop after one cycle of the external field. The dashed curve is the
hysteresis loop after the second cycle. 
Another sweep of the external field would retrace
the solid curve, indicating that this particular realization of
randomness undergoes a 2-cycle. $B_e^{max}$ is less than
the saturation field $B_e^{sat}.$ (If
$B_e^{max}$ were increased beyond $B_e^{sat}$, no multicycles
would be found.)}  
\label{fig1}
\end{center}
\end{figure}

Figure~\ref{fig2} shows the major hysteresis loop for a sample
realization of randomness.  Since all the pillars are magnetized
vertically each time $B_e = \pm B_e^{max} = \pm \infty,$ $m$ is
trivially equal to 1. However, one can see the avalanching dynamics
characteristic of this system, and the fact --- discussed earlier
--- that the two halves of the major loop are not complementary.
Figure~\ref{fig1} shows a hysteresis 2-cycle.
 
\section{Results}
Using an algorithm that performs the operations of the previous
section, we search through a large number of realizations of
randomness to find regions in parameter space where the probability
of finding multicycles is high.  The parameters in the model are
radius $R$ and height $h$ of the pillars, $B_e^{max}$, and the
damping coefficient $\gamma_2$.  If the lattice spacing, $R$ and
$h$ are all scaled by a factor $\lambda,$ all the terms in the
Hamiltonian are scaled by $\lambda^3,$ which does not affect the
ratios of the terms. Accordingly, the lattice spacing can be set
to 100 nm without loss of generality.  Given a set of parameters,
the algorithm determines the periodicity $m$ for a given realization
of randomness.  By classifying the periodicity for a large number
of realizations, we obtained the approximate probability for finding
an $m$-cycle as a function of $m$.

The easiest parameter to vary experimentally is $B_e^{max}$.  As
mentioned in the previous section, when $B_e^{max}$ is too large
or too small, multicycles will not be present.  We find that
multicycles can be roughly optimally found when $B_e^{max}$ approaches
the saturation field $B_e^{sat}$ but not greater.  
$B_e^{sat}$ is
different for every realization of randomness, therefore the optimal
field can only be determined by scanning over various values of
$B_e^{max}$.  The range of $B_e^{max}$ where the occurrence of
multicycles is appreciable depends on the pillar dimensions.  For
$R$=30 nm and $h$=180 nm, the probability of finding multicycles
when $B_e^{max}$ is optimal is $\sim 2-3$ times the probability
when $B_e^{max}$ is $\sim$15 percent from the optimal field.  In
general, for systems in which the probability of finding multicycles
is small, the range of $B_e^{max}$ is narrow.  The narrow range in
$B_e^{max}$ is not an obstacle to finding multicycles, due to the
ease of tuning this parameter experimentally.

The damping coefficient of the LLG equation, $\gamma_2$, cannot be
easily calculated.  In fact, different experimental environments
could allow for a large range of $\gamma_2$.  Because of our inability
to obtain a reasonable and realistic approximation for $\gamma_2$,
we run searches for a wide range of $\gamma_2$ (relative to
$\gamma_1$).  The results show that multicycles exist for very small
$\gamma_2$ to essentially infinite damping.  (The large damping limit
is implemented by setting $\gamma_1$=0 and keeping $\gamma_2$ finite.)
Small values of $\gamma_2$ tend to give more multicycles, as one
might expect: the probability of finding multicycles increases by
a factor of $\sim 1.5$ when $\gamma_2$ is reduced from $\gamma_1$
to $\sim 0.1 \gamma_1.$ Unexpectedly, the multicycle probability
also seems to increase slightly when $\gamma_2$ is larger than
$\gamma_1$.  When the dynamical equation is strictly dissipative
($\gamma_1$=0), the multicycle probability is comparable to the
probability when $\gamma_2 \gtrsim \gamma_1$. There are no clear
trends in the distribution of $m$ when $\gamma_2$ is changed.

\begin{table}
\caption{\label{table2} Approximate probabilities of finding an
$m$-cycle for systems with pillars of different radii.  The lattice
spacing is 100 nm and the aspect ratio, $h/R$, is 5. }
\begin{ruledtabular} 
\begin{tabular}{cccccc}
 R (nm) & $P_{m=2}$ & $P_{m=3}$  & $P_{m=4}$ & $P_{m>4}$ & 
$P_{m>1}$\\
\hline
10 & 0 & 0 & 0 & 0 & 0 \\ 
20 & .08 & .02 & 0 & 0 & .1 \\
25 & .14 & .12 & .02 & .04 & .32 \\
30 & .2 & .06 & .04 & .16 & .46 \\
35 & .22 & .22 & .04 & .12 & .6 \\
40 & .22 & .12 & .1 & .18 & .62 \\
45 & .28 & .12 & 0 & .10 & .50 
\end{tabular} 
\end{ruledtabular} 
\end{table}

Because of the difficulty in calculating $\gamma_2$, we conservatively
set $\gamma_2$ to a value where the probability of finding multicycles
is approximately minimal.  As mentioned in the preceding paragraph,
this occurs when $\gamma_2 \approx \gamma_1$ .  With $B_e^{max}$ at
its optimal value, and $\gamma_2 = \gamma_1 = 1$, probabilities
of finding multicycles for systems with pillars of different $R$
and $h$ are found.  The different terms in the Hamiltonian scale
differently as $R$ and $h$ are varied.  Table~\ref{table2} shows
results for systems of the same aspect ratio ($h/R = 5$) and different
radii.  The maximum radius for a pillar is 45 nm, with the lattice
spacing 100 nm.  From Table~\ref{table2}, one can see that systems
of pillars with larger radii generally display multicycles more often.

The occurrence of multicycles depends largely on the aspect ratio of
the pillars.  Avalanches tend to occur only for pillars where $h/R$
is large; accordingly, systems with disk-shaped pillars, {\it i.e.\/}
with negative $d_z,$ do not display multicycles.  In fact, even
when $d_z$ is positive, multicycles are only found for a sufficiently
large shape anisotropy energy.  Table~\ref{table3} shows numerical
results for systems of pillars with a 30 nm radius and various
aspect ratios.  Multicycles are only likely to be found for systems
with long pillars.  No multicycles are found for disk-shaped pillars
as expected.

\begin{table}
\caption{\label{table3} Approximate probabilities of finding
an $m$-cycle for systems with pillars of different aspect ratios
$h/R$.  The lattice spacing is 100 nm and the radius is fixed at
30 nm.   }
\begin{ruledtabular}
\begin{tabular}{cccccc}
$h/R$ & $P_{m=2}$ & $P_{m=3}$  & $P_{m=4}$ & $P_{m>4}$ & $P_{m>1}$\\
\hline
.5 & 0 & 0 & 0 & 0 & 0 \\
1 & 0 & 0 & 0 & 0 & 0 \\
3 & .16 & .04 & .02 & .12 & .34 \\
4 & .12 & .08 & .02 & .16 & .38 \\
5 & .2 & .06 & .04 & .16 & .46 \\
6 & .16 & .1 & .06 & .16 & .48 \\
10 & .24 & .2 & .06 & .1 & .6 
\end{tabular}
\end{ruledtabular}
\end{table}

Systems with pillar vacancies are also studied.  Vacancies are
implemented by introducing a finite probability for a pillar to be
missing at every lattice site.  One might expect that these
vacancies would introduce more randomness in the system, thereby
increasing the number of multicycles.  For pillars with R=30nm and
h=150nm, the probability of finding a multicycle is $\sim 0.46$
without any vacancies.  When the probability of having a vacancy
at a site is small ($\sim 0.2$), the number of multicycles drops
by about 40 percent.  When the vacancy probability increases to
$\gtrsim 0.5$, the probability of finding multicycles decreases to
less than 0.1.  This decrease in probability could be due to the
decrease in number of pillars.  We conclude that, contrary to what
one might expect,  random vacancies
do not increase the probability of multicycles .

If a square lattice is used instead of a triangular one, the dipolar
couplings between nearest neighbor pillars are not frustrated. In
a checkerboard pattern, all nearest neighbor bonds would be satisfied,
but the next nearest neighbor bonds (and certain neighbors further 
apart) would not be satisfied. We find 
that the amount of frustration in the square lattice is sufficient
for multicycles to be found, though the probability of finding a
multicycle is significantly less on a square lattice than on a
triangular lattice.  The number of multicycles found on a square
lattice is approximately half of the number found on a triangular
lattice.

A possible mechanism for increasing frustration and disorder is to
introduce random ferromagnetic couplings between pillars.  These
bonds could be manufactured by building one large pillar instead
of two small ones.  The large pillar would have two domains that
are coupled ferromagnetically.  By randomly assigning the bonds
with some probability, more disorder can be introduced in addition
to that from the crystal anisotropy.
We did not study how these ferromagnetic bonds affect the number
of multicycles.

One interesting question is whether exact Return Point
Memory~\cite{Sethna} survives when it is extended beyond the random
field Ising model with purely ferromagnetic interactions to continuous
time vector models. To answer that, we use the LLG model with nearest
neighbor random ferromagnetic interactions and the same crystalline and shape
anisotropy model that is used above (without dipole coupling).
We apply a random field and
search for violations of RPM with different random seeds. We start at high
fields and go to a minimum field of $-1.7$ and record the spin
configuration. Then we go up to a field $1.7$ and back down to $-1.7$.
We find that for a $4\times4$ square lattice of spins, that of order
1 percent of systems {\em violate} RPM because the
initial and final minimum configurations and total magnetization are
substantially different. Violation of RPM is seen both
with and without the precessional term in Eq.(\ref{LLG}).  This shows
that it is not possible to extend the proof of RPM to continuous
vector models.

\section{Conclusions}
In this paper, we have investigated the feasibility of observing multicycles
and non-complementary hysteresis loops in a candidate experimental
system: that of cylindrical magnetic nano-pillars arranged on a
lattice.  We have performed realistic numerical simulations of this
system, by calculating the magnetic interactions between the pillars
and then employing continuous spin dynamics and the Landau Lifshitz
Gilbert equation to obtain their time evolution. Using physically
appropriate parameters, we have shown that there is often multicycle
hysteretic behavior, i.e. a periodic adiabatic external magnetic
field causes a subharmonic steady state response in the magnetization.
We have also shown that, even though the Hamiltonian is invariant
under spin and field reversal, the dynamics are not, so that the
two branches of the major hysteresis loop are not complementary.
This result cannot be obtained with an Ising approximation to the
system and its dynamics.

Further implications of the non-complementary nature of LLG dynamics
will be studied in future research. 

Because systems of this kind are currently the subject of much
experimental investigation, we believe that it would be fruitful
to attempt to observe the unusual behavior predicted here.

\section{Acknowledgments }
We thank John Donohue, Holger Schmidt, Mark Sherwin, and Larry Sorensen for very
useful discussions.


\end{document}